%% file: main.tex
\RequirePackage{fix-cm}
\documentclass[twocolumn]{svjour3}       
\smartqed  

\usepackage[T1]{fontenc}
\usepackage{graphicx}
\usepackage{todonotes}

\usepackage{multirow}
\usepackage{booktabs}

\newcommand*\ON[0]{$\surd$}
\usepackage{url}
\usepackage{pgfplots}
\usepackage{filecontents}
\usepackage{float}
\usepackage{longtable}
\usepackage{multicol}
\usepackage{longtable}

\usepackage{tabularx}
\newcolumntype{b}{X}
\newcolumntype{s}{>{\hsize=.4\hsize}X}
\newcolumntype{m}{>{\hsize=.7\hsize}X}
\newcolumntype{P}[1]{>{\raggedright\arraybackslash}p{#1}}
\usepackage{listings}
\lstdefinestyle{mystyle}{
    keepspaces=true,                 
    numbers=left,                    
    numbersep=5pt,                  
    showspaces=false,                
    showstringspaces=false,
    showtabs=false
}
\begin{document}

\title{Security Assurance Cases - State of the Art of an Emerging Approach}



\author{Mazen Mohamad         \and
        Jan-Philipp Stegh\"ofer \and
        Riccardo Scandariato
}

\institute{Mazen Mohamad \at
              Department of Computer Science and Engineering, University of Gothenburg $\vert$ Chalmers University of Technology, Gothenburg SE-41296, Sweden \\
              \email{mazen.mohamad@gu.se}           
           \and
           Jan-Philipp Stegh\"ofer \at
              Department of Computer Science and Engineering, University of Gothenburg $\vert$ Chalmers University of Technology, Gothenburg SE-41296, Sweden \\
              \email{jan-philipp.steghofer@cse.gu.se}
              \and
           Riccardo Scandariato \at
              Department of Computer Science and Engineering, University of Gothenburg $\vert$ Chalmers University of Technology, Gothenburg SE-41296, Sweden \\
              \email{riccardo.scandariato@cse.gu.se}
              }

\date{Received: date / Accepted: date}

\maketitle

\begin{abstract}
Security Assurance Cases (SAC) are a form of structured argumentation used to reason about the security properties of a system. After the successful adoption of assurance cases for safety, SAC are getting significant traction in recent years, especially in safety-critical industries (e.g., automotive), where there is an increasing pressure to be compliant with several security standards and regulations.
Accordingly, research in the field of SAC has flourished in the past decade, with different approaches being investigated.
In an effort to systematize this active field of research, we conducted a systematic literature review (SLR) of the existing academic studies on SAC. 
Our review resulted in the in-depth analysis and comparison of 51 papers.
Our results indicate that, while there are numerous papers discussing the importance of security assurance cases and their usage scenarios, the literature is still immature with respect to concrete support for practitioners on how to build and maintain a SAC. More importantly, even though some methodologies are available, their validation and tool support is still lacking.
\end{abstract}

\input{Introduction.tex}
\input{RelatedWork.tex}
\input{ResearchMethod.tex}
\input{Results.tex}

\input{Discussion.tex}

\input{ThreatsToValidity.tex}
\input{Conclusion.tex}

\begin{acknowledgements}
This work is partially supported by the CASUS research project funded by VINNOVA, a Swedish funding agency.
\end{acknowledgements}


\bibliographystyle{spmpsci}
\bibliography{references}

%
%


%
%

\end{document}

%% file: Introduction.tex
\section{Introduction}
\label{sec:intro}

A security assurance case (a.k.a. security case, or SAC) is a structured set of arguments that are supported by material evidence and can be used to reason about the the security posture of a software system.
Security assurance cases represent an emerging trend in the secure development of critical systems, especially in domains like automotive and healthcare.
The adoption of security cases in these industries is compelled by the recent introduction of standards and legislation. 
For instance, the upcoming standard ISO/SAE 21434 on Road Vehicles Cybersecurity Engineering includes the explicit requirement to create `cybersecurity cases' to show that a vehicle's computing infrastructure is secure.

The creation of a security case, however, is far from trivial, especially for large organizations with complex product development structures.
For instance, some technical choices about the security case might require a change of the development process.
For example, the security case shown in Figure \ref{fig:sacExample} (and discussed in Section \ref{sec:relWork}) requires that a thorough threat analysis is conducted throughout the product structure and at different stages of the development. 
If this is not the case, a thorough re-organization of the way of working is necessary, or maybe the security case should have been structured in a different shape. 
Also, the construction of a security case often requires the collaboration of several stakeholders in the organization, e.g., to ensure that all the necessary evidence is collected from the software and process artifacts.

To summarize, companies are facing the conundrum of making both urgent and challenging decisions concerning the adoption of security cases.
Naturally, they could refer to academic research, which has published a relatively large number of papers on the topic in recent years.
In order to facilitate such endeavor, this paper presents a systematic literature review (SLR) of research papers on security cases.
To the best of our knowledge, this is the first study of this kind in this field.
This SLR collects most relevant resources (51 papers) and presents their analysis according to a rich set of attributes like, the types of augmentation structures that are proposed in the literature (threat identification --used in Figure \ref{fig:sacExample}-- being one option), the maturity of the existing approaches, the ease of adoption, the availability of tool support, and so on.
Ultimately, this paper presents a complete guideline to adoption geared towards practitioners.
To this aim, we have created a workflow describing the suggested activities that are involved in the adoption process for security cases.
Each stage of the workflow is annotated with a suggested reading list, which refers to the papers included in this SLR. 
We remark that the SLR also represents a useful tool for academics in order identify research gaps and opportunities, which are discussed in this paper as well.

The rest of the paper is structured as follows. 
In Section \ref{sec:relWork} we provide some background on assurance cases and discuss the related work.
In Section \ref{sec:resmethod} we describe the research questions and the methodology of this study.
In Section \ref{sec:res} we list the papers included in this study and present the results of the analysis.
In Section \ref{sec:disc} we further discuss the results and present the workflow for security cases adoption.
Finally, Section \ref{sec:con} presents the concluding remarks.

%% file: RelatedWork.tex
\section{Background and Related Work}
\label{sec:relWork}

\subsection{Assurance cases}

\begin{figure}[tb]
\begin{center}
\includegraphics[width=\columnwidth]{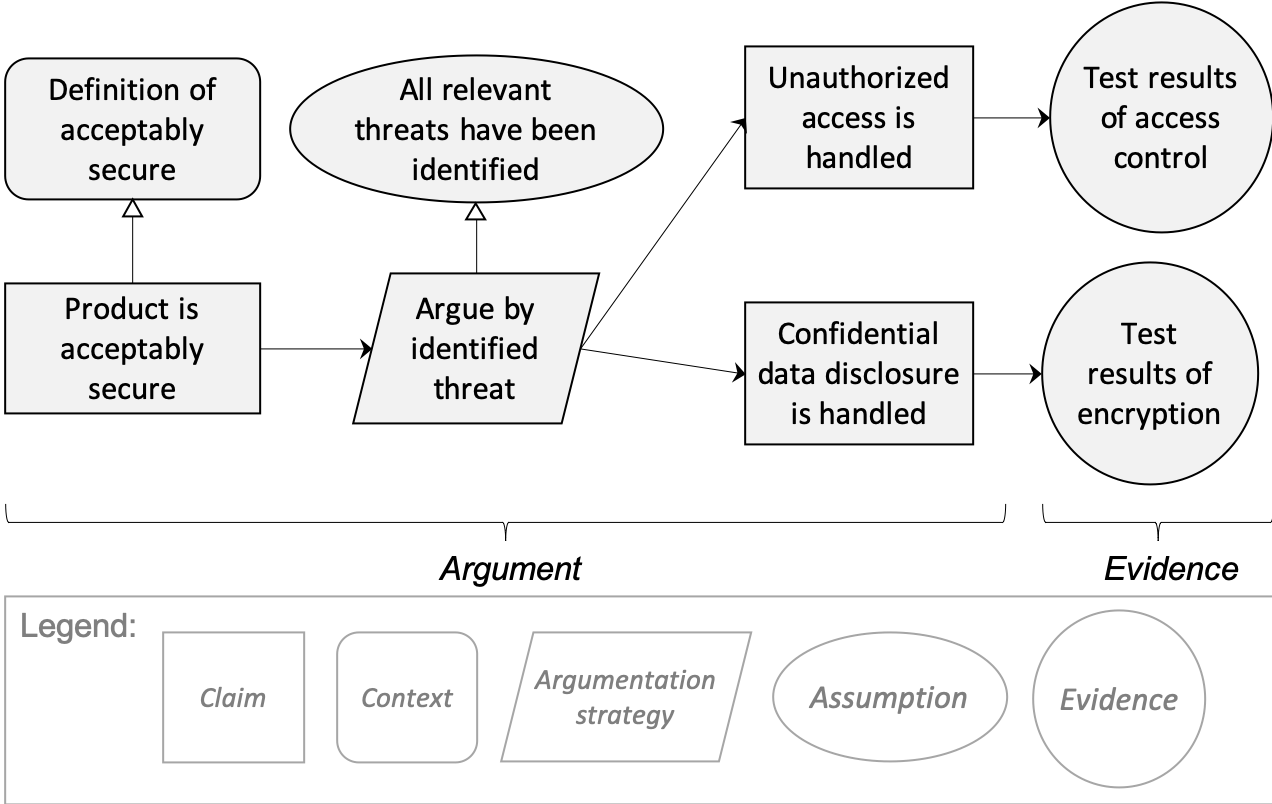}
\end{center}
\caption{An example of an assurance case} 
\label{fig:sacExample}
\end{figure}

Assurance cases are defined by the GSN standard \cite{GSN_standard} as \emph{``A reasoned and compelling argument, supported by a body of evidence, that a system, service or organisation will operate as intended for a defined application in a defined environment.''}. 
Assurance cases can be documented in either textual or graphical forms.
Figure \ref{fig:sacExample} depicts a very simple example of an assurance case and its two main parts, i.e., the argument and the evidence.
The case in the figure follows the GSN notation \cite{gsn}, and consists of the following nodes: claim (also called goal), context, strategy, assumption (also called justification), and evidence (also called solution). 
At the top of the case, there is usually a high level claim, which is broken down to sub-claims based on certain strategies.
The claims specifies the goals we want to assure in the case, e.g., that a certain system is secure.
An example of a strategy is to break down a claim based on different security attributes.
The breaking down of claims is repeatedly done until we reach a point where evidence can be assigned to justify the claims/sub-claims. 
Examples of evidence are test results, monitoring reports, and code review reports. 
The assumptions made while applying the strategies, e.g., that all relevant threats have been identified, are made explicit using the assumption nodes. 
The context of the claims is also explicitly set in the context nodes. An example of a context is the definition of an acceptably secure system. 

Assurance cases have been widely used for safety critical systems in multiple domains \cite{bloomfield2010}. An example is the automotive industry, where safety cases have been used for demonstrating compliance with the functional safety standard ISO 26262 \cite{palin2011,birch2013,iso26262}. 
However, there is an increasing interest in using these cases for security as well. 
For instance, the upcoming automotive standard ISO 23434 \cite{iso21434} explicitly requires the creation of cyber-security arguments. 
Security assurance cases are special types of assurance cases where the claims are about the security of the system in question, and the body of evidence justifies the security claims. 

\subsection{Related work}
To the best of our knowledge this study is the first systematic literature review on security assurance cases. However, there have been studies covering the literature on safety assurance cases.

Nair et al. \cite{nair2013} conducted a systematic literature review to classify artefacts which can be considered as safety evidence. The researchers contributed with a taxonomy of the evidence, and listed the most frequent evidence types referred to in literature. The results of the study show that the structure of safety evidence is mostly induced by the argumentation and that the assessment of the evidence is done in a qualitative manner in the majority of cases in contrast to quantitative assessment. Finally, the researchers list eight challenges related to safety evidence, and the creation of safety cases was the second most mentioned one in literature according to the study. 

Maksimov et al. \cite{maksimov2018} contributed with a systematic literature review of assurance case tools. The researchers list 37 tools that have been developed in the past two decades, and an analysis of their functionalities. The study also includes an evaluation of the reported tools on multiple aspects, such as creation support, maintenance, assessment, and reporting. In our study, we also review supporting tools for assurance cases' creation, but we focus on the reported tools specifically for SAC.

Gade et al. \cite{gade2015} conducted a literature review of assurance-driven software design. The researchers provide a review of 15 research papers
with an explanation of the techniques and methodologies each of these papers provide with regards to assurance-driven software design. This work intersects with our work in that assurance-driven software design can be used as a methodology or approach for creating assurance cases. However, unlike Gade et al. \cite{gade2015} our study focuses on SAC, and is done in a systematic way.



%% file: ResearchMethod.tex
\section{Research Method}
\label{sec:resmethod}

We conducted a systematic literature review following the guidelines introduced by Kitchenham et al.~\cite{kh2007guidelines}.


\subsection{Research questions and assessment criteria}
\label{sec:resmethod:rq}


This study aims at answering the following research questions.

\begin{table}[tb]
\caption{Assessment criteria for RQ1 (rationale).}
\label{method_charrq1}
\begin{tabularx}{\columnwidth}{bb}
\toprule
\textbf{RQ1 criteria} & Values \\
\midrule
Motivation     & E.g., compliance to standards \\ 
               & \dots \\ 
\midrule
Usage scenario & E.g., support for court case \\ 
               & \dots \\ 
\bottomrule
\end{tabularx}
\end{table}

[\textbf{RQ1}] \textbf{RATIONALE --- In the literature, what rationale is provided  to support the adoption of SAC?} 
In particular, we are interested in whether there are statements that go beyond the intuitive rationale of using SAC ``for security assurance''. 
For instance, our initial research \cite{industrialNeeds} indicated that compliance with security standards and regulations is also an important driver. 
As shown in Table \ref{method_charrq1}, to answer this research question we analyze the surveyed papers and extract two characteristics:
\begin{enumerate}
\item[(i)] 
\emph{Motivation}, i.e., the reason for using SAC as stated by the researchers. We used two criteria for determining whether a certain study provides a motivation for using SAC. That is, the wording has to be explicit (i.e., there must be a reference to usage or advantage) and specific (i.e., providing some details). 
\item[(ii)] 
\emph{Usage scenario}, i.e., scenarios in which SAC could be used to achieve additional goals, next to to security assurance. We used the same criteria (explicit and specific mention) used for the motivation.
\end{enumerate}

\begin{table}[tb]
\caption{Assessment criteria for RQ2 (construction)}
\label{method_charrq2}
\begin{tabularx}{\columnwidth}{bb}
\toprule
\textbf{RQ2 criteria } & Values \\ 
\midrule
Approach provided & Yes / No \\ 
\midrule
Coverage & Argumentation \\
         & Evidence     \\ 
         & Generic (i.e., both) \\
\midrule
Argumentation & E.g., based on threat avoidance \\
(if covered)  & \dots \\
\midrule
Evidence     & E.g., collect test results \\
(if covered) & \dots \\
\bottomrule
\end{tabularx}
\end{table}

[\textbf{RQ2}] \textbf{CONSTRUCTION --- In the literature, which approaches are reported for the construction of security assurance cases, and which aspects do the approaches cover?} 
This question aims at inventorying the existing approaches for creating SAC, which is a challenging task for adopters. As shown in Table \ref{method_charrq2}, we also assess the \emph{coverage} of the approach, i.e., whether it can be used for creating the argumentation, for collecting the evidence, or both. 
Finally, for each covered part of the SAC, we identify the gist of the approach with respect to the suggested \emph{argumentation} strategy and the types of \emph{evidence} to be used in creating SACs.

\begin{table}[tb]
\caption{Assessment criteria for RQ3 (support)}
\label{method_charrq3}
\begin{tabularx}{\columnwidth}{bbb}
\midrule
\textbf{RQ3 criteria } &  & Values \\
\midrule
Prerequisites &                 & E.g., threat modeling is performed \\
              &                 & \dots \\
\midrule
Patterns      &                 & E.g., a catalog or argumentation patterns is provided \\
              &                 & \dots \\
\midrule
Tool support  & Tool mentioned  & Yes / No        \\
              & Type of tool    & Created  / Used \\
\midrule
Notation      &                 & Graphical (GSN)  \\
              &                 & Graphical (CAE)  \\
              &                 & Textual          \\
              &                 & \dots            \\
\bottomrule
\end{tabularx}
\end{table}

[\textbf{RQ3}] \textbf{SUPPORT --- In the literature, what practical support is offered to facilitate the adoption of SAC?}
The purpose of this question is to understand the practicalities of creating and working with SAC.
With reference to Table \ref{method_charrq3}, first we study the approaches and identify the conditions (i.e., \emph{prerequisites}) that have to be met in order for the outcome of the paper to be applicable. These prerequisites 
Second, we check whether the papers propose libraries of \emph{patterns} or templeteized SAC, as these are extremely usefult for non-expert adopters.
Third, we analyze the \emph{tool support}. We check whether the paper suggests the usage of a tool for any of the activities related to SAC. In case it does, we extract the description of that tool, and whether it was created by the researchers or if it is a third party tool used in the paper.
The last characteristic in this research question is the \emph{notation} used to represent the SAC. The most common ones are GSN~\cite{gsn}, and CAE~\cite{CAE}, but there are other notations such as plain text.

\begin{table}
\caption{Assessment criteria for RQ4 (validation).}
\label{method_charrq4}
\begin{tabularx}{\columnwidth}{bm}
\toprule
\textbf{RQ4 criteria} & Values \\
\midrule
Type & Illustration \\
     & Case study \\
     & Experiment \\
     & Other \\
\midrule
Domain & Medical \\
       & Automotive \\
       & Software engineering \\
       & \dots \\
\midrule
Data source & Research project \\
            & Commercial product \\
            & \dots \\
\midrule
SAC created & Yes / No   \\
\midrule
Creators   & Academic authors \\
           & Industrial authors \\
           & 3rd party experts \\
\midrule
Validators & Academic authors \\
           & Industrial authors \\
           & 3rd party experts \\
\bottomrule
\end{tabularx}
\end{table}

[\textbf{RQ4}] \textbf{VALIDATION --- In the literature, what evidence is provided concerning the validity of the reported approaches?} 
Our interest is to understand how the approaches and usage scenarios of SAC are validated (or supported by any evidence). 
We aim at identifying: 
\begin{itemize}
    \item[i] The \emph{type} of the validation conducted in the study, e.g., case study, or experiment. Note that `case study' is a widely used term to refer to worked examples \cite{caseStudies,runeson2009}. In this work, we consider a validation conducted in an industrial context to be a case study \cite{yin2003}, and those done within a research context to be illustrations.  Experiments are studies in which independent variables are manipulated to test their effect on dependent variables \cite{caseStudies}.
    \item[ii] The \emph{domain} (i.e., application area) in which the validation is conducted.
    \item[iii] The \emph{source} of the data used for the validation, e.g., a research project or a commercial product.
    \item[iv] Whether or not a SAC is \emph{created} as part of the validation process. 
    \item[v] In case a SAC is created, we look for its \emph{creators}. This characteristic has three levels, which are academic authors, authors with industrial background, or expert group.
    \item[vi] The \emph{validators}, i.e., the parties that conducted the validation (similar to the creators).
\end{itemize}


\subsection{Performing the systematic review}

We performed a search for papers related to SAC by means of 3 scientific search engines: IEEE Xplore, ACM Digital Library, and Elsevier Scopus. 
We did not use Google Scholar as, in our own prior experience, the results from this search engine overlap with the results of the engines we we mentioned above.

\subsubsection{Constructing the search string}

To maximize the chance of obtaining all the relevant papers in the field, the search string used in the search engines must contain keywords that are commonly used in said papers.
Therefore, prior to constructing the search string, we familiarized ourselves with the specific terminology used by researchers in the field of security assurance cases.
To do so, we conducted a manual search for papers related to SAC that were published in the past five years in the following venues: SAFECOMP, CCS, SecDev, ESSOS, ISSRE, ARES, S\&P, Asia CCS, and ESORICS.
The selection of the venues was based on their high visibility in the security domain.

Next, we created the search string. 
In particular, we used two groups of keywords. 
The first group (line 1 below) is meant to scope the area of the study, while the second group (lines 2--4) included the terms referring to the parts of an assurance case. As a result, the we formed the search string as follows:

\begin{lstlisting}[language=Python]
( security OR privacy OR trust ) AND 
( claim OR argument OR evidence 
  OR justification OR 'assurance case' 
  OR assurance )
\end{lstlisting}

As a quality check for our search string, we used three relevant known studies \cite{8_finnegan2014,10_othmane2014,20_xu2017}, which are listed in IEEE Xplore.
We ran the query in the same library and confirmed that those papers were returned.

\subsubsection{Inclusion and exclusion criteria}

\begin{table}
\caption{Inclusion and exclusion criteria.}
\label{method_inex}
\begin{tabular}{l}
\toprule
\textbf{Inclusion criteria} \\ 
\midrule
1. Studies addressing the creation, management, \\ 
   or application of security assurance cases. \\
2. Studies related to security/privacy/trust assurance. \\
3. Studies related to security/privacy/trust argumentation. \\
\midrule
\textbf{Exclusion criteria}\\ 
\midrule
1.  Studies written in any language other than English. \\
2. Studies published before 2004. \\
3. Short papers (less than 3 pages). \\
4. Studies focusing on risk/threat/hazard detection. \\
5. Studies addressing risk/threat/hazard analysis. \\
6. Studies addressing cryptography. \\
7. Studies focusing on security assessment/evaluation. \\
8. Studies about (only) safety assurance. \\
\bottomrule
\end{tabular}
\end{table}

The inclusion and exclusion criteria are shown in table \ref{method_inex}.
The inclusion criteria are rather straightforward, considering to the nature of this SLR.
Concerning the exclusion criteria, we have decided to only consider studies written in English language, as this is the common language among the authors of this SLR. 
Further, SAC have been the focus of research only in recent times (although assurance cases, in general, have been around for much longer) and the field is rapidly evolving. Hence, we restricted our SLR to the past 15 years, also to to avoid outdated results.
We also excluded short papers, as answering our research questions requires studies with results rather than only ideas.
Finally, exclusion criteria 4--8 exclude studies that focus on topics that are marginally related to SAC but would not not help us answer our research questions.

\subsubsection{Searching and filtering the results}

We executed the query on three libraries (IEEE Xplore, ACM Digital Library, and Scopus) in January 2019, and got the results shown in Table \ref{method_res}.
In the case of Scopus, we limited the search to the domains of either computer science or engineering. Also, because of the high number of returned results from Scopus, we decided to limit the included studies to the first 2000 after ordering the results based on relevance. We believe that the considered studies were sufficient, as the last 200 papers of the retained set from Scopus (i.e., papers 1801-2000) were all excluded when we applied the first filtering round (see below).

\begin{table}[tb]
		\caption{Number of included studies after each round of filtration.}
		\label{method_res}
		\begin{tabularx}{\columnwidth}{lllll}
		\toprule
	        & & \multicolumn{3}{l}{\textbf{After filtering round}} \\
	        \textbf{Library} & \textbf{Papers} & \textbf{1st}  & \textbf{2nd} & \textbf{3rd} \\
        \midrule
			IEEE Xplore & 4513 & 118 & 23 & 22 \\
			ACM DL      & 1927 &  35 &  3 &  3 \\
			Scopus      & 2000 &  68 & 23 & 19 \\

			            &      &     &    &\textbf{+7} (snowballing) \\
		\midrule
			\textbf{Total}& \textbf{8440} &   &   &\textbf{51} \\
		\bottomrule
		\end{tabularx}
\end{table}

In the first filtering round, we applied the inclusion and exclusion criteria to the titles and keywords of all the results we got from the search (8440 papers). As shown in Table \ref{method_res} This round reduced the number of studies to 211 papers. 
In the second filtering round, we applied the inclusion and exclusion criteria\footnote{Except for exclusion criteria 1,2, and 8, which only needed to be applied once.} to the abstracts and conclusions of the 211 remaining studies. After this step, the number of studies was reduced to 49.
In the last filtering round, we fully read the remaining 49 papers, applied the inclusion and exclusion criteria on the whole text, and ended up with 44 included studies.

After finishing the three filtering rounds, we started performing the \emph{snowball search} method \cite{snowballing}. This step added seven papers. Hence, the total number of included studies was raised to \textbf{51 papers}.

%% file: Results.tex
\section{Results}
\label{sec:res}

In this section, we provide a descriptive analysis of the included papers in this SLR, and then present the results and answers to our four research questions.

\subsection{Descriptive statistics}
Figure \ref{fig:M1} shows the years when our 51 included studies were published. The graph shows a peak of 10 publications in 2015, which indicates an increase in interest in the research filed compared to previous years, especially the time between 2005 and 2012 where the number of publications was three or less each year.

\begin{figure}
	\centering
\begin{tikzpicture}
        \begin{axis}[
            symbolic x coords={Conference, Workshop, Journal, Other},
            xtick=data,
            xlabel={Type of publication},
			ylabel={Number of publications},
            grid style=dashed
          ]
            \addplot[ybar,fill=gray] coordinates {
                (Conference,   18)
                (Workshop,  17)
                (Journal,   13)
                (Other, 3)
            };
        \end{axis}
\end{tikzpicture}
\caption{Types of publication of the included studies } 
\label{fig:M2}
\end{figure}
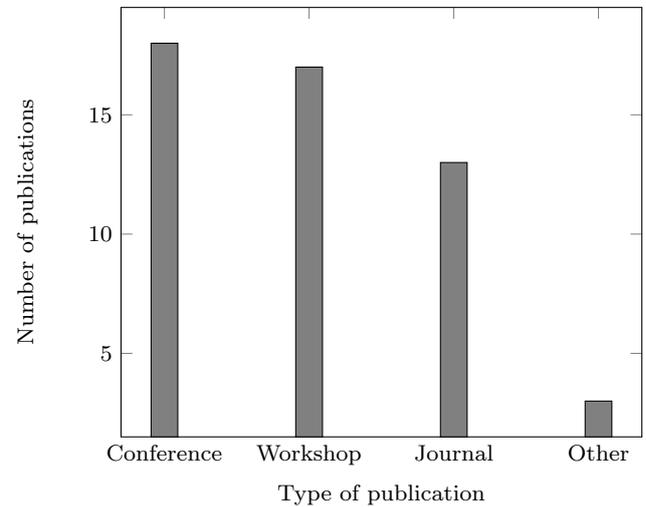

Figure \ref{fig:M2} shows the venues where the included studies were published. The graph shows that most of the publications were made in conferences and workshops (18 and 17 respectively). 13 of the papers were published in journals, and three were technical reports.
We also looked into the authors of the selected papers to find the portion of the papers with at least one author from industry. We found that less than 25\% (12 papers)~\cite{2_cockram2007,3_goodger2012,14_netkachova2015,19_netkachova2016,20_xu2017,24_gacek2014,25_rodes2014,29_bloomfield2017,30_netkachova2014,37_gallo2015,39_cheah2018,47_ionita2017} included at least one author from industry.
\subsection{RQ1: Motivation}
In order to find the rational reported in literature for the adoption of SAC, we looked into motivations and usage scenarios, as explained in~\ref{sec:resmethod:rq}. 
Some of the identified motivations in RQ1.1 could also be seen as usage scenario. For example, \emph{compliance with standards and regulation} could be seen as a motivation for using SAC, but also as a purpose for which SAC could be used.  

\subsubsection{Motivation}
Generally speaking, the main reason that motivates usage of SAC is to perform security assurance on a system. In this study, we looked for other motivations, and more specific ones. We extracted this piece of information by looking for motivations to use SAC rather than the suggested approach or method for creating them. In some of the papers, the motivation was made explicit in a separate section, or as the focus of the whole study (e.g. \cite{50_knight2015,53_alexander2011}). However, in most papers, this was discussed either in the introduction and background sections, or as a part of motivating the used or suggested approach for creating SAC. If a study discusses only the basic SAC benefits, or is not being specific about the motivation (e.g. motivates security assurance in general), then we have categorised this paper as one that did not discuss any motivations for using SAC.
\\
Table \ref{res_rq1_tab_motive} shows all motivations found in the 51 studies included in this paper.
The results show that about 73\% of the studies included at least one motivation for using SAC other than security assurance. The most common motivations are: \emph{(i)} compliance with standards and regulation (8); \emph{(ii)} it is a proven approach from the safety domain (6); and \emph{(iii)} compliance with requirements (4).\\
Categorizing the motivations resulted in the following categories:
\begin{itemize}
    \item External forces: Compliance with standards and regulation, and compliance with requirements (in case of suppliers).
    \item Process improvement: SAC helps in integrating security assurance with the development process. Moreover, they help factoring work per work items, and analyzing complex systems. 
    \item Structure and documentation: The structure of SAC implies a way of work that reduces technical risks, and enhances security communication among stakeholders. 
    \item Security assessment: SAC help in assessing security and spotting weaknesses in security for the systems in question. Hence, they help building confidence in the those systems.
    \item Knowledge transfer: It is a proven approach in safety which has been used effectively for a long time, and could be similarly in security.
\end{itemize}
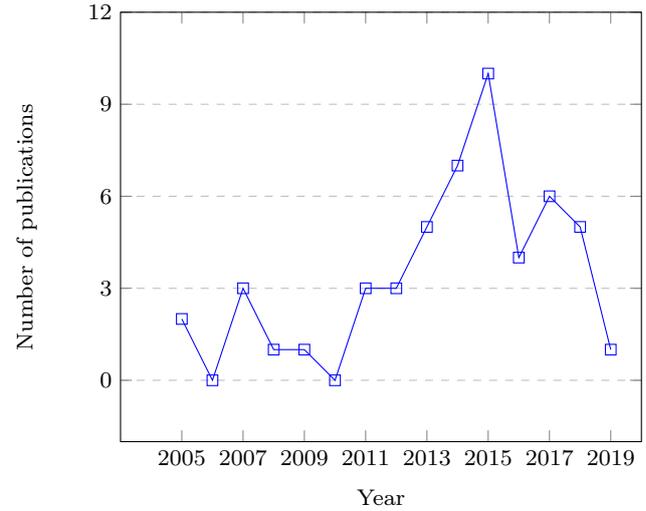
\begin{figure}
	\centering
	\begin{tikzpicture}
		\begin{axis}[
				title={},
				xlabel={Year},
				ylabel={Number of publications},
				xmin=2003, xmax=2020,
				ymin=-2, ymax=12,
				xtick={2005,2007,2009,2011,2013,2015,2017,2019},
				ytick={0,3,6,9,12},
				legend pos=north west,
				ymajorgrids=true,
				grid style=dashed,
			/pgf/number format/.cd,
            use comma,
            1000 sep={}]
			\addplot[
				color=blue,
				mark=square,
			]
			coordinates {
				(2005,2)(2006,0)((2007,3)(2008,1)(2009,1)(2010,0)(2011,3)(2012,3)(2013,5)(2014,7)(2015,10)(2016,4)(2017,6)(2018,5)(2019,1)
			};
		\end{axis}
	\end{tikzpicture}
	\caption{Publication year of the included studies} \label{fig:M1}
\end{figure}

\begin{table*}
	\caption{RQ1 --- Papers stating the motivations for using SAC}
	\label{res_rq1_tab_motive}
	\noindent\makebox[\textwidth]{%
		\begin{tabular}{@{}llP{8.5cm}@{}}
		\toprule
			\textbf{Study}
			
			& \textbf{Category} 
			& \textbf{Motivation}\\
			\midrule
			                      

			                       \textbf{Ankrum et al.} \cite{9_ankum2005} & External forces & Comply with standards and regulation
			                      \\ \textbf{Calinescu et al.} \cite{55_calinescu2017}  & External forces  & Comply with security requirements of safety-critical systems
			                      \\ \textbf{Cyra et al.} \cite{1_cyra2007}  &External forces & Comply with standards and regulation
			                      \\ \textbf{Finnegan et al.} \cite{8_finnegan2014} & External forces  & Comply with regulation and maintain confidence in the product in question
			                      \\ \textbf{Finnegan et al. (2)} \cite{45_finnegan2014} & External forces  & Comply with regulation 
			                      \\ \textbf{He et al.} \cite{4_he2012} & External forces & Reason about cybersecurity policies and procedures
			                      \\ \textbf{Mohammadi et al.} \cite{51_mohammadi2018} &  External forces &  Learn from the safety domain where it is a proven approach
			                      \\ \textbf{Ray et al.} \cite{16_ray2015} & External forces & Comply with regulation and internal needs from cyber-physical systems' manufacturers
			                      \\ \textbf{Sklyar et al. (2)} \cite{21_sklyar2017,28_sklyar2017,31_sklyar2019} & External forces & Comply with standards
			                      \\ \textbf{Sljivo et al.} \cite{23_sljivo2016} & External forces & Comply with standards and regulation
			                      \\ \textbf{Strielkina et al.} \cite{22_strielkina2018} & External forces& Comply with security regulation
			                      \\ 
			                       \textbf{Goodger et al.} \cite{3_goodger2012} & Knowledge transfer & Learn from the safety domain to integrate oversight for safety and security
			                      \\ \textbf{Ionita et al.} \cite{47_ionita2017} & Knowledge transfer & Learn from the safety domain where it is a proven approach
			                      \\ \textbf{Netkachova et al. (2)} \cite{30_netkachova2014} & Knowledge transfer & Learn from the safety domain where it is a proven approach
			                      \\ \textbf{Poreddy et al.} \cite{26_poreddy2011} & Knowledge transfer& Learn from the safety domain, where it is a proven approach
			                      \\ \textbf{Sklyar et al.} \cite{17_sklyar2016} & Knowledge transfer & Learn from the safety domain, where it is a proven in-use approach
			                      \\ 
			                       \textbf{Ben Othmane et al.} \cite{18_othmane2016} & Process improvement & Trace security requirements and assure security during iterative development.
			                      \\ \textbf{Ben Othmane et al.} \cite{10_othmane2014} & Process improvement & Assure security during iterative development
			                      \\ \textbf{Cheah et al.} \cite{39_cheah2018}  & Process improvement & Cope with the increasing connectivity of systems 
			                      \\ \textbf{Cockram et al.} \cite{2_cockram2007} & Process improvement & Reduces both technical and program risks through process improvement
			                      \\ \textbf{Gallo et al.} \cite{37_gallo2015}  & Process improvement & Factor analytical and implementation work per component, requisite, technology, or life-cycle
			                      \\ \textbf{Lipson et al.} \cite{56_lipson2008}& Process improvement & Help analyzing complex systems
			                      \\ \textbf{Netkachova et al.} \cite{19_netkachova2016} & Process improvement & Tackle security issues which have intensified challenges of engineering safety-critical systems.
			                      \\ \textbf{Weinstock et al.} \cite{52_weinstock2007} & Process improvement & Include people and processes in security assurance in addition to technology
			                      \\ 
			                       \textbf{Alexander et al.} \cite{53_alexander2011} & Security assessment & Help security evaluators to focus their attention on critical parts of the system
			                       \\ \textbf{Bloomfield et al.} \cite{29_bloomfield2017} & Security assessment & Ensure the fulfillment of security requirements
			                       \\ \textbf{Finnegan et al.} \cite{45_finnegan2014} &Security assessment & Improve overall security practices and demonstrate confidence in security
			                      \\ \textbf{Hawkins et al.} \cite{15_hawkins2015} & Security assessment & Justify and assess confidence in critical properties
			                      \\ \textbf{Knight} \cite{50_knight2015} &Security assessment& Spot security related weaknesses in the system
			                      \\ \textbf{Poreddy et al.} \cite{26_poreddy2011} &Security assessment&  Assist in identifying security loopholes while changing the system
			                      \\ \textbf{Rodes et al.} \cite{25_rodes2014} & Security assessment & Measure software security
			                      \\ \textbf{Strielkina et al.} \cite{22_strielkina2018} & Security assessment & Acquire an input for decision making of requirement conformity 
			                      \\ \textbf{Vivas et al.} \cite{34_vivas2011} & Security assessment & Acquire confidence that the security of the system meets the requirements
			                      \\ 
			                      \textbf{Agudo et al.} \cite{38_agudo2009}  & Structure \& documentation & Incorporate certifications and evaluation methods in an evidence-based structure
			                      \\ \textbf{Alexander et al.} \cite{53_alexander2011} &Structure \& documentation &   Summarize security thinking when vendors are involved
			                      \\ \textbf{Finnegan et al.} \cite{43_finnegan2013}  & Structure \& documentation & Communicate and report achieved security level
			                      \\ \textbf{Knight} \cite{50_knight2015} & Structure \& documentation& Document rational for security claims
			                      \\ \textbf{Netkachova et al.} \cite{14_netkachova2015} & Structure \& documentation & Aid in communication as it provides a summary of issues and their interrelationship
			                      \\ \textbf{Patu et al.} \cite{5_patu2013} & Structure \& documentation & Aid in the survival of modern system, with respect to security challenges
			                      \\ \textbf{Ray et al.} \cite{16_ray2015} &Structure \& documentation& Comply with internal needs from cyber-physical systems' manufacturers
			\\ \bottomrule
		\end{tabular}}
\end{table*}

\subsubsection{Usage Scenarios}
While SACs are usually used to establish evidence-based security assurance for a given system, researchers have reported cases where SAC could be used to achieve different goals. We looked into studies that focus on using SAC for a purpose other than security assurance, or that is specific to a certain domain (e.g. security assurance for medical devices), or context (e.g. security assurance within the agile framework). 
Table \ref{res_rq1_tab_us} shows the usage scenarios of SAC found in literature. We were able to extract usage scenarios from 14 different papers (28\% of the total number of papers).  The usage scenarios found show a wide range of applications of SAC. Seven of the papers suggest using SAC for evaluating different parts of the system or its surroundings. This includes evaluating the system architecture \cite{49_yamamoto2015}, security in a specific context \cite{30_netkachova2014}, confidence in security \cite{25_rodes2014}, trustworthiness of the system \cite{51_mohammadi2018}, validation of service grade \cite{6_masumoto2013}, satisfaction of requirements \cite{54_haley2005}, and security standards \cite{13_graydon2015}. The remaining seven papers suggest using SAC for process improvement \cite{38_agudo2009}, development of security features \cite{18_othmane2016,10_othmane2014}, development of security policies and strategies \cite{29_bloomfield2017}, asset management \cite{3_goodger2012}, and teaching information security \cite{37_gallo2015}.

\begin{table*}
	\caption{RQ1 --- Papers relevant to understanding the usage scenarios}
	\label{res_rq1_tab_us}
		\begin{tabular}{ll}
			\toprule
			\textbf{Study}
			& \textbf{Usage scenarios}
            
			\\ \midrule
			                       \textbf{Agudo et al.} \cite{38_agudo2009}    & Integrating security engineering and assurance based development using SAC    
			                      \\ \textbf{Ben Othmane et al.} \cite{18_othmane2016} &  Controlling the impact of incremental development on security assurance using SAC
			                      \\ \textbf{Ben Othmane et al. (2)} \cite{10_othmane2014} & Developing iterative security features using SAC for security assurance  
			                      \\ \textbf{Bloomfield et al.} \cite{29_bloomfield2017}   & Using SAC in the development of security strategies and policies 
			                      \\ \textbf{Finnegan et al.} \cite{43_finnegan2013}  & Reporting the achieved security level using SAC
			                      \\ \textbf{Gallo et al.} \cite{37_gallo2015}   & Using SAC to teach information security
			                      \\ \textbf{Goodger et al.} \cite{3_goodger2012}  & Asset management
			                      \\ \textbf{Graydon et al.} \cite{46_graydon2013}  & Evaluation of security standards
			                      \\ \textbf{Haley et al.} \cite{54_haley2005} & Using SAC to prove achieving satisfaction of security requirements 
			                      \\ \textbf{Masumoto et al.} \cite{6_masumoto2013}  & SAC is used to validate that a service satisfies a certain service grade
			                      \\ \textbf{Mohammadi et al.} \cite{51_mohammadi2018}  & Ensuring trustworthiness in cyber-physical systems using trustworthiness cases.
			                      \\ \textbf{Netkachova et al. (2)} \cite{30_netkachova2014}  & Evaluation of security of critical infrastructures using SAC
			                      \\ \textbf{Rodes et al.} \cite{25_rodes2014}  & Measuring software security based on confidence in security argument
			                      \\ \textbf{Yamamoto} \cite{49_yamamoto2015}  &  Evaluation of system architecture based on security  claims 
			\\ \bottomrule
		\end{tabular}
\end{table*}

\subsection{RQ2: Approaches}
We were able to find 26 different approaches in the literature. These studies focus on creating either complete security assurance cases, or parts of them (argumentation, or evidence).
Table \ref{res_rq2_tab_approaches} shows the found approaches, which part/s of SAC they cover, argumentation strategies used to divide the claims and create the arguments, and the evidence used to justify the claims in the approaches.
We categorize the approaches as follows:
\begin{itemize}
    \item Integrating SAC in the development life-cycle: These approaches suggest mapping the SAC creation activities to the development activities to integrate SACs in the development and security processes \cite{38_agudo2009,10_othmane2014,16_ray2015,34_vivas2011}, as well as assurance case driven design \cite{17_sklyar2016,21_sklyar2017,28_sklyar2017,31_sklyar2019}.
    \item Using different types of AC for security: These approaches suggest using different types of assurance cases other than SAC for security assurance. These types are: trust cases \cite{1_cyra2007}, trustworthiness cases \cite{42_gorski2012,51_mohammadi2018}, combined safety and security cases \cite{2_cockram2007},dynamic assurance cases \cite{55_calinescu2017},multiple viewpoint assurance cases where security is treated as an assurance viewpoint \cite{23_sljivo2016}, and dependability cases \cite{41_patu2013}.
    \item Documenting and visualizing SAC: These studies give guidelines of how to document a SAC, and visualize it \cite{26_poreddy2011,40_coffey2014,52_weinstock2007}. In this category there are papers that focus on a specific part of SAC. These are:
    \subitem Argumentation-centric: These approaches focus on the argumentation part of the SACs.  Different strategies are suggested in literature: security standards-based argument \cite{1_cyra2007,8_finnegan2014,9_ankum2005}, and satisfaction argument \cite{54_haley2005}. Structures of argumentation found in literature are: model-based \cite{15_hawkins2015}, and layered structure \cite{14_netkachova2015,20_xu2017}. Moreover, we have one study which suggests an automatic creation of argument graphs \cite{7_tippenhauer2014}. 
    \subitem Evidence-centric: These approaches focus mainly on different aspects of SACs' evidence. These aspects are: searching for evidence \cite{11_chindamaikul2014}, collecting and generating evidence \cite{36_shortt2015,56_lipson2008}, and rating of potential artifacts to be used as evidence \cite{39_cheah2018}.
    
  
\end{itemize}


\subsubsection{Coverage}
As shown in Table~\ref{res_rq2_tab_approaches}, 16 of the found approaches cover the creation of complete SACs, six focus on argumentation, and the remaining four on evidence.
Five out of the 16 studies to create complete security cases did not include any examples of evidence used to justify the claims.
\subsubsection{Argumentation}
Argumentation is a very important part of SAC, and forms the bigger part of it. The argumentation starts with a security claim, and continues as the claim is being broken down into sub-claims. The strategy is used to provide a means by which the breaking down of claims is being done. Each level of the argumentation could be done with a different strategy than the other levels. Hence, one SAC might have one or more argumentation strategies, which is the case in some of the included studies in this SLR.
We looked for an explicit mention of the used strategy. If none was provided, we looked into the example cases to find the used argumentation strategy. 
Here, we look at the argumentation strategies used in the different approaches. We could not find any correlation between the approach, and the argumentation strategy used in the approach. For instance, the approaches which integrate SAC within the development life-cycle can have different argumentation strategies, e.g., requirements~\cite{38_agudo2009} and development phases~\cite{16_ray2015}.
The most common strategy depends on the output of a threat, vulnerability, asset or risk analysis (8 papers). Other popular strategies are: breaking down the claims based on the requirements, or more specifically quality requirements and even more specifically security requirements (5 papers), and arguing based on security properties (5 papers). Additionally, researchers also used system and security goals (4 papers), software components or features (3 papers), security standards and principles (2 papers), pre-defined argumentation model (1 paper), and development life-cycle phases (1 paper).    
Table \ref{res_rq2_tab_approaches} shows the approaches we found in literature with the respective argumentation strategies used in each of them.

\subsubsection{Evidence}
Even though evidence is a very important part of SAC, and a complex one as well, only four of 26 included approaches focused on it. Even in the generic approaches, there was a much deeper focus on the argumentation than the evidence. This explains why five out of the 16 generic approaches did not even include an example of what evidence would look like in their approach.
We found evidence either by looking for explicitly mentioned ones in the articles, or by extracting the evidence part from the reported SACs.
Cheah et al.~\cite{39_cheah2018} present a classification of security test results using security severity ratings. This classification can be included in the security evaluation, which may be used to improve the selection of evidence when creating security assurance cases. Chindamaikul et al.~\cite{11_chindamaikul2014} investigate how information retrieval techniques, and formal concept analysis can be used to find security evidence in a document corpus. 
Shrott and Weber \cite{36_shortt2015} present a method to apply fuzz testing to support the creation of evidence for security assurance cases. Lipson and Weinstock \cite{56_lipson2008} describe how to understand, gather, and generate multiple kinds of evidence that can contribute to building SAC.
The most common types of evidence reported in literature are testing results (12 papers)~\cite{18_othmane2016,55_calinescu2017,39_cheah2018,11_chindamaikul2014,56_lipson2008,26_poreddy2011,36_shortt2015,17_sklyar2016,21_sklyar2017,28_sklyar2017,31_sklyar2019,23_sljivo2016}, and different types of analysis. These analysis include threat and vulnerability \cite{2_cockram2007,8_finnegan2014,45_finnegan2014,41_patu2013}, code and bug \cite{11_chindamaikul2014,18_othmane2016,17_sklyar2016,21_sklyar2017,28_sklyar2017,31_sklyar2019}, security standards and policies \cite{38_agudo2009,14_netkachova2015}, risk \cite{51_mohammadi2018}, and log analysis \cite{51_mohammadi2018,41_patu2013}. Other types of evidence reported in literature include process documents \cite{56_lipson2008}, design techniques \cite{51_mohammadi2018}, and security awareness and training \cite{41_patu2013,56_lipson2008}.
Table \ref{res_rq2_tab_approaches} shows the approaches we found in literature with the respective evidence types used in each of them.


\begin{table*}
	\caption{RQ2 --- Papers presenting approaches to construct SAC}
	\label{res_rq2_tab_approaches}
	\begin{tabular}{@{}P{2.8cm}P{3.6cm}lP{4cm}P{3.8cm}@{}}
			\toprule
			\textbf{Study}
			& \textbf{Approach}
			& \textbf{Coverage}
			& \textbf{Argumentation}
			& \textbf{Evidence}
			\\ \midrule
			                       \textbf{Agudo et al.} \cite{38_agudo2009}   & Assurance-based development & Generic    & Requirements, system goals, system views and models & Common criteria evaluation 
			                      \\ \textbf{Ankrum et al.} \cite{9_ankum2005}   & Mapping SAC to standards & Generic & Security standard description & NA 
			                      \\ \textbf{Ben Othmane et al.}  \cite{10_othmane2014} & Security assurance for incremental software development & Generic & Security goals & Test results, code review
			                      \\ \textbf{Calinescu et al.}  \cite{55_calinescu2017}   &Dynamic assurance cases & Generic &  Requirements   & Verification results 
			                      \\ \textbf{Cheah et al.} \cite{39_cheah2018}   & Systematic Security Evaluation & Evidence & NA & Rated test results 
			                      \\ \textbf{Chindamaikul et al.} \cite{11_chindamaikul2014}  & Document retrieval and formal concept analysis & Evidence &  Security properties   & Test results, bug fixes reports
			                      \\ \textbf{Cockram et al.} \cite{2_cockram2007} &Dependability by contract & Argument & Vulnerabilities, threats, and mitigation & Threat analysis report, vulnerability analysis report 
			                      \\ \textbf{Coffey et al.} \cite{40_coffey2014}   & Concept map-based & Generic & Vulnerabilities & NA 
			                      \\ \textbf{Cyra et al.} \cite{1_cyra2007}  & Using Trust-cases to comply with security standards & Argument & Risks & NA
			                      \\ \textbf{Finnegan et al.} \cite{8_finnegan2014,45_finnegan2014}  & Risk based approach & Generic & Security capabilities, mitigation controls & Threat logs, vulnerability logs 
			                      \\ \textbf{Go\`rski et al.} \cite{42_gorski2012}  & TRUST-IT Maintenance and assessment of trustworthiness arguments & Generic & Toulmin's argument~\cite{toulmin2003}& NA 
			                      \\ \textbf{Haley et al.} \cite{54_haley2005} & Satisfaction arguments & Argument & Security requirements & NA 
			                      \\ \textbf{Hawkins et al.} \cite{15_hawkins2015} & Model-based assurance & Argument & Software components & NA 
			                      \\ \textbf{Lipson et al.} \cite{56_lipson2008}& Evidence-based assurance of security properties & Evidence & NA & Process documents, test results, and many more.
			                      \\ \textbf{Mohammadi et al.} \cite{51_mohammadi2018} & Trustworthiness cases & Generic & Availability, threat analysis, goals satisfaction & Risk assessment reports, log analysis, design techniques
			                      \\ \textbf{Netkachova et al.} \cite{14_netkachova2015} & Layered Approach & Generic & Source of security requirements, changes during life-cycle & Security analysis, security policies
			                      \\ \textbf{Patu et al. (2)} \cite{41_patu2013} & Evidence-based dependability case & Generic & Vulnerabilities & Countermeasures to vulnerabilities, security awareness documents, access control, logs 
			                      \\ \textbf{Poreddy et al.} \cite{26_poreddy2011} & Documenting AC for Security & Generic & Security properties & Test results 
			                      \\ \textbf{Ray et al.} \cite{16_ray2015} & Integrating security engineering and AC development & Generic & Development life-cycle phases & NA 
			                      \\ \textbf{Shortt et al.} \cite{36_shortt2015} & Hermes Targeted fuzz testing & Evidence & NA & Test results 
			                      \\ \textbf{Sklyar et al.} \cite{17_sklyar2016,28_sklyar2017,21_sklyar2017,31_sklyar2019} & Assurance Case Driven Design & Generic & Quality requirements, security properties, features, components, software layers, green IT principles & Code analysis, testing reports  
			                      \\ \textbf{Sljivo et al.} \cite{23_sljivo2016} & Multiple-viewpoint AC &Generic & Contracts (pair of assumptions and guarantees) & Test results 
			                      \\ \textbf{Tippenhauer et al.} \cite{7_tippenhauer2014} & Automatic generation of security argument graphs & Argument & Security goals  & NA 
			                      \\ \textbf{Vivas et al.} \cite{34_vivas2011} & Security assurance driven software development & Generic & Threats, vulnerabilities & NA 
			                      \\ \textbf{Weinstock et al.} \cite{52_weinstock2007} & Arguing security & Generic & Prevention and detection & Test results, Analysis tool results, developers' training 
			                      \\ \textbf{Xu et al.} \cite{20_xu2017} & Layered Argument strategy & Argument & Assets, threats & NA 
			\\ \bottomrule
		\end{tabular}
\end{table*}

\subsection{RQ3: Support}

\subsubsection{Tools:}
We found 16 software tools which have been used one way or another in the creation on security assurance cases in literature. 
Seven out of the found tools were created by the researchers. Four of these seven target assurance cases in general~\cite{35_fung2018,24_gacek2014,15_hawkins2015,7_tippenhauer2014}, while the remaining three are created to be used in the creation of SAC specifically~\cite{18_othmane2016,39_cheah2018,36_shortt2015}.
Table~\ref{res_rq3_tab_tools} shows the found tools, and the respective studies in which the tools are used. A brief description of the main functionalities of the tools, as well as whether the tools are created or used by the authors are also presented.
There are four main types of reported tools: 
\begin{itemize}
    \item Creation tools: These are used to create and document assurance cases in general\cite{tool_adrelard,tool_turboac,24_gacek2014,d-case}
    \item Argumentation tools: These focus mainly on the creation of the argumentation part of SAC\cite{15_hawkins2015,tool_openArgue,tool_arg_sec,7_tippenhauer2014}
    \item Evidence tools: These focus on the creation of SAC evidence\cite{tool_uppaal,36_shortt2015}
    \item Support tools: These are used to assist the creators of SAC in the analysis needed for creating SAC\cite{tool_meld,18_othmane2016,39_cheah2018}, maintenance of SAC\cite{35_fung2018}, and management of different parts of SAC\cite{tool_norsta}.
\end{itemize}

In addition to these software tools, a concept called concept lattice was used as a tool to help users determine the relevance of a given document to be used as an evidence\cite{11_chindamaikul2014}.

\begin{table*}
	\caption{RQ3 --- Tools supporting the creation, documentation, and visualization of SAC.}
	\label{res_rq3_tab_tools}
	\begin{tabular}{@{}lP{2.5cm}P{6.8cm}l@{}}
			\toprule
			\textbf{Study}
			& \textbf{Tool support}
			& \textbf{Description}
			& \textbf{Used or created}
			
			\\ \midrule
                                 \textbf{Ankrum et al.}~\cite{9_ankum2005}&Adelard Safety Case Editor (ASCE)~\cite{tool_adrelard} & Notation tool used for creating SACs &Used
                                \\ \textbf{Ben Othmane et al.}~\cite{18_othmane2016}&Meld\cite{tool_meld}&  Visual diff and merge tool &Used
                                \\  &SECUREAGILE &  Tracing the impact of code changes on security & Created
                                \\ \textbf{Calinescu et al.}~\cite{55_calinescu2017} &UPPAAL~\cite{tool_uppaal}& A verification tool which helps produce assurance evidence & Used
                                \\ \textbf{Cheah et al.}~\cite{39_cheah2018} &Software tool (no specific name)&Semi-automated test execution tool & Created
                                \\ \textbf{Finnegan et al.}~\cite{43_finnegan2013}&TurboAC~\cite{tool_turboac} &Generating and documenting SACs& Used 
                                \\ \textbf{Fung et al.}~\cite{35_fung2018}&MMINT-A & Automated change impact assessment on assurance cases& Created
                                \\ \textbf{Gacek et al.}~\cite{24_gacek2014}&Resolute & Constructing assurance cases based on architecture analysis and design language models & Created
                                \\ \textbf{Go\`rski et al.}~\cite{42_gorski2012}&NOR-STA~\cite{tool_norsta}& A set of services used for argumentation editing and assessment, as well as evidence repositories& Used
                                \\ \textbf{Hawkins et al.}~\cite{15_hawkins2015}&Instantiation program (no specific name)&Creates an instantiation model, which is then converted into a GSN argument & Created
                                \\ \textbf{Ionita et al.}~\cite{47_ionita2017}&OpenArgue~\cite{tool_openArgue}  & The study compares these argument modelling tools & Used 
                                \\ & &Argumentation Sheets
                                \\ & &ArgueSecure~\cite{tool_arg_sec}
                                \\ \textbf{Patu et al.}~\cite{41_patu2013}& D-Case Editor~\cite{d-case} & Assurance case editor and a GSN Pattern Library for Eclipse & Used
                                \\ \textbf{Poreddy et al.}~\cite{26_poreddy2011}&Adelard Safety Case Editor (ASCE)~\cite{tool_adrelard} & Notation tool used for creating SACs& Used
                                \\ \textbf{Shortt et al.}~\cite{36_shortt2015} &Hermes& Dynamic code coverage analysis for evidence creation &Created
                                \\ \textbf{Tippenhauer et al.}~\cite{7_tippenhauer2014}&CyberSAGE~\cite{tool_CyberSAGE}& Automatically generates argument graphs & Created
                        	\\ \bottomrule
	\end{tabular}
\end{table*}


\subsubsection{Prerequisites:}
Prerequisites are the conditions that need to be met in order for the outcome of a study to work or be applied. We found prerequisites in the included studies by checking the inputs of the proposed outcomes (approaches, usage scenarios, tools, and patterns). If an input is not a part of the outcome itself, we considered it to be a prerequisite to that outcome. 
Table \ref{res_rq3_tab_prereq} shows the found prerequisites, and the respective type of study for each. There are 17 reported prerequisites. The majority belong to approaches (11) and the remaining belong to usage scenarios (3), patterns (2), and tools (1).
We categorize the prerequisites in four categories as follows:
\begin{itemize}
    \item Usage of specific format~\cite{24_gacek2014,15_hawkins2015,23_sljivo2016}
    \item Existence of analysis and modelling~\cite{39_cheah2018,3_goodger2012,5_patu2013,20_xu2017}
    \item Usage of specific documents and repositories~\cite{11_chindamaikul2014,2_cockram2007,4_he2012,41_patu2013,7_tippenhauer2014,34_vivas2011}, and security standards~\cite{9_ankum2005,46_graydon2013,1_cyra2007}
    \item Existence of special expertise~\cite{29_bloomfield2017}
\end{itemize}

\begin{table*}
	\caption{RQ3 --- Papers discussing the prerequisites of SAC approaches, usage scenarios, and tools.}
	\label{res_rq3_tab_prereq}
	\begin{tabular}{@{}llP{9.6cm}@{}}
			\toprule
			\textbf{Study}
			& \textbf{Type of study}
			& \textbf{Prerequisites}
			\\ \midrule
                                 \textbf{Ankrum et al.}  \cite{9_ankum2005}&Approach&Security standards
                                \\ \textbf{Bloomfield et al.}  \cite{29_bloomfield2017}&Usage scenario& Special expertise (regulator in this case)
                                \\ \textbf{Cheah et al.} \cite{39_cheah2018} &Approach&Threat model
                                \\ \textbf{Chindamaikul et al.}   \cite{11_chindamaikul2014}&Approach &Evidence corpus
                                \\ \textbf{Cockram et al.}  \cite{2_cockram2007}&Approach&Module boundary contract
                                \\ \textbf{Cyra et al.}    \cite{1_cyra2007}&Approach&Security standards
                                \\ \textbf{Gacek et al.}   \cite{24_gacek2014}&Tool&Use of AADL (Architecture Analysis and Design Language)
                                \\ \textbf{Goodger et al.}  \cite{3_goodger2012}&Usage Scenario&Asset and vulnerability analysis
                                \\ \textbf{Graydon et al.}  \cite{46_graydon2013}&Usage Scenario& Security standards
                                \\ \textbf{Hawkins et al.}  \cite{15_hawkins2015}&Approach&Argument patterns, and models containing needed information for instantiation
                                \\ \textbf{He et al.}  \cite{4_he2012}&SAC Pattern&Lessons learned
                                \\ \textbf{Patu et al.} \cite{5_patu2013}&SAC patterns&Asset analysis
                                \\ \textbf{Patu et al. (2)}  \cite{41_patu2013}&Approach&A pre-defined list of common risks/vulnerabilities and solutions in the domain
                                \\ \textbf{Sljivo et al.}  \cite{23_sljivo2016}&Approach&Contracts in the extended SEooCMM format (a special format)
                                \\ \textbf{Tippenhauer et al.}  \cite{7_tippenhauer2014}&Approach&Graph extension templates, and sub-graphs derived from security assessment
                                \\ \textbf{Vivas et al.}  \cite{34_vivas2011}&Approach&Well defined SDLC (Software Development Life-Cycle) process
                                \\ \textbf{Xu et al.}  \cite{20_xu2017}&Approach&Threat model, and asset model
            \\ \bottomrule
	\end{tabular}
\end{table*}

\subsubsection{Patterns} Patterns of SAC exist, as there are reoccurring claims and arguments. Using patterns can save the creators of SACs a lot of time and effort. We found ten studies which deal with patterns. Six of these create their own argumentation patterns~\cite{8_finnegan2014,45_finnegan2014,4_he2012,5_patu2013,26_poreddy2011,20_xu2017}.
The remaining four include usage of patterns~\cite{15_hawkins2015,7_tippenhauer2014}, a guideline for creating and documenting security case patterns \cite{52_weinstock2007}, and a catalogue of security and safety case patterns \cite{48_taguchi2014}.
In some studies, the examples are taken from the safety domain, and there is a usage of safety patterns e.g.~\cite{55_calinescu2017}. However, in this study, we only considered patterns created and used for security assurance cases.
Table \ref{res_rq3_tab_patterns} shows the papers that deal with security assurance case patterns.

\begin{table*}
	\caption{RQ3 --- Papers presenting patterns.}
	\label{res_rq3_tab_patterns}
	\begin{tabular}{@{}lP{11.7cm}@{}}
			\toprule\textbf{Study}
			& \textbf{Description of the pattern-based approach}
			\\ \midrule
                                 \textbf{Finnegan et al.}  \cite{8_finnegan2014,45_finnegan2014} &Creation of security capability argument pattern which takes a risk based approach, and argues for each security capability defined in a technical report for risk management in medical devices.
                                \\ \textbf{Hawkins et al.}  \cite{15_hawkins2015}&Usage of argument patterns as input to the model-based approach for building assurance case arguments. A suggested pattern argues over individual software components.
                                \\ \textbf{He et al.} \cite{4_he2012}&Creation of generic cases which use security arguments that are informed by security incidents in healthcare organizations. 
                                \\ \textbf{Patu et al.}  \cite{5_patu2013}&Creation of security patterns at the requirement phase of the system development life-cycle. One suggested pattern argues over security attributes.
                                \\ \textbf{Poreddy et al.}  \cite{26_poreddy2011}&Creation of assurance case patterns, suggested argumentation strategies are: integrity, availability, reliability, confidentiality and maintainability.
                                \\ \textbf{Taguchi et al.}  \cite{48_taguchi2014}&A catalogue of safety and security case patterns. The patterns are derived from process patterns through a literature survey.  
                                \\ \textbf{Tippenhauer et al.}  \cite{7_tippenhauer2014}&Usage of argument patterns to automatically generate argument graphs. The paper includes five different patterns categorized into two categories: inter-type, and intra-type.
                                \\ \textbf{Weinstock et al.}  \cite{52_weinstock2007}&A guideline of how to create and use security assurance case patterns is presented. An example pattern is also presented.
                                \\ \textbf{Xu et al.}  \cite{20_xu2017}& Creation of different argument patterns to be used in different layers to form a layered argument structure.
            \\ \bottomrule
	\end{tabular}
\end{table*}

\subsubsection{Notations}
41 out of our 51 included papers specify at least one notation to be used for expressing and documenting a SAC. The most common notation is the Goal Structure Notation (GSN)~\cite{gsn} which is suggested by 26 studies. Another popular notation is the Claim Argument Evidence (CAE)~\cite{CAE} notation which is suggested by nine studies. Other notation are: text (6 studies), concept maps~\cite{40_coffey2014} (1), and Claim-Argument-Evidence Criteria (CAEC)~\cite{19_netkachova2016,14_netkachova2015,30_netkachova2014} notation which is extension of the CAE notation (3 studies of the same authors).

\subsection{RQ4: Validation}
We consider validation to be the process to confirm the applicability of an outcome in the selected studies as validation. In case validation is performed in a selected study, we looked for the type of validation, the domain of application, the source of data, whether a SAC is created during the validation, the creators of the SACs, and who performed the validation. Table \ref{res_rq4_tab_validation} shows the results of RQ4.
As shown in the table, 36 studies include a validation of the outcome. The majority of the outcomes (21) were validated using illustrative cases, 11 were validated using case studies, and the remaining four used experiments (3) and observation as a part of an Action Design Research (ADR)~\cite{ADR} study. 
The data sources vary among the validations, as can be seen in Table \ref{res_rq4_tab_validation}. We categorize these sources into three main categories:
\begin{itemize}
    \item Research, open source, and in-house projects (20) \cite{9_ankum2005,11_chindamaikul2014,2_cockram2007,40_coffey2014,24_gacek2014,54_haley2005,15_hawkins2015,51_mohammadi2018,14_netkachova2015,41_patu2013,26_poreddy2011,16_ray2015,25_rodes2014,36_shortt2015,31_sklyar2019,23_sljivo2016,22_strielkina2018,7_tippenhauer2014,34_vivas2011,37_gallo2015}
    \item Commercial products / systems (9) \cite{18_othmane2016,10_othmane2014,55_calinescu2017,39_cheah2018,3_goodger2012,42_gorski2012,6_masumoto2013,20_xu2017,30_netkachova2014}
    \item Standards, regulation, and technical reports (7) \cite{29_bloomfield2017,1_cyra2007,45_finnegan2014,35_fung2018,46_graydon2013,4_he2012,28_sklyar2017}
\end{itemize}


SACs were presented in 31 out of the 36 validations. Representing a complete SAC is in most cases not possible even in small illustration cases, because of the  amount of information required to build one. However, the scale of SAC representation in the included validations varies to a large extent. Some validations present an example of a full branch of SAC, i.e., a claim all the way from top to evidence (e.g., He and Johnson~\cite{4_he2012}), while others present very brief examples of SACs (e.g., Gallo and Dahab~\cite{37_gallo2015}). Table~\ref{res_rq4_tab_validation} also shows the creators of the SACs in each study. In only two cases, experts were used to create the SACs. In the majority of the studies (28), the authors created the SACs. However, eight of these had authors from industry. These are shown in Table \ref{res_rq4_tab_validation} as Authors* in the Creators column. 

Table~\ref{res_rq4_tab_validation} also shows the domains in which the validation was conducted. The most common domains are: Software Engineering (7), and Medical (7).

The last column in Table \ref{res_rq4_tab_validation} shows the persons which performed the validation in each study. Out of the 36 included validations, only five used 3rd parties to validate the outcomes. In the remaining 31 validations, the authors performed the validation. However, eight out of these had authors from industry. These are shown in Table \ref{res_rq4_tab_validation} as Authors* in the Validator column.

\begin{table*}
	\caption{RQ4 --- Papers presenting a form of validation}
	\label{res_rq4_tab_validation}
	\noindent\makebox[\textwidth]{%
	\begin{tabular}{@{}llP{1.8cm}P{4.2cm}lll@{}}
			\toprule
			\textbf{Study}
			&\textbf{Validation}
			& \textbf{Domain}
			& \textbf{Data source}
			& \textbf{SAC}
			& \textbf{Creators}
			& \textbf{Validators}
			
			\\ \midrule
                                 \textbf{Ankrum et al.} \cite{9_ankum2005}&Illustration&Security&Research security project&&Authors&Authors
                                \\ \textbf{Ben Othmane et al.} \cite{18_othmane2016}&Case study&Software Engineering&E-Commerce product&\ON&Authors&3rd party
                                \\ \textbf{Ben Othmane et al. (2)} \cite{10_othmane2014}&Illustration&Telecom&Commercial project&\ON&Authors&Authors
                                \\ \textbf{Bloomfield et al.}  \cite{29_bloomfield2017}&Case study&Safety&Regulatory organization&\ON&Authors*&3rd party
                                \\ \textbf{Calinescu et al.} \cite{55_calinescu2017} &Case study&Marine, Trading&Underwater Vehicle System, Trading System&\ON&Authors&Authors
                                \\ \textbf{Cheah et al.} \cite{39_cheah2018} &Case study&Automotive&Vehicle infotainment system, diagnostics tool&\ON&Authers*&3rd party
                                \\ \textbf{Chindamaikul et al.}   \cite{11_chindamaikul2014}&Experiment&Information Retrieval&Open source software development project&\ON&Expert group&Authors
                                \\ \textbf{Cockram et al.}  \cite{2_cockram2007}&Illustration&SafSec&Command and control system for locating persons&\ON&Authors*&Authors*
                                \\ \textbf{Coffey et al.}  \cite{40_coffey2014}&Illustration&Software Engineering&SOA composite application&\ON&Expert group&3rd party
                                \\ \textbf{Cyra et al.}    \cite{1_cyra2007}&Illustration&Security&Security standard BS 7799-2 &\ON&Authors&Authors
                                \\ \textbf{Finnegan et al. (2)}   \cite{45_finnegan2014}&Observation&Medical&Technical report&\ON&Authors&3rd party
                                \\ \textbf{Fung et al.} \cite{35_fung2018}&Case study&Automotive&Power sliding door -- Case from ISO26262 standard&\ON&Authors&Authors
                                \\ \textbf{Gacek et al.}  \cite{24_gacek2014}&Illustration&Embedded Systems&Research project for unmanned air vehicles&\ON&Authors*&Authors*
                                \\ \textbf{Gallo et al.}   \cite{37_gallo2015}&Experiment&Education&Course in information security&\ON&NA&Authors*
                                \\ \textbf{Goodger et al.} \cite{3_goodger2012}&Case study&Critical infrastructure&Critical information Infrastructure&&NA&Authors*
                                \\ \textbf{G\`orski et al.}    \cite{42_gorski2012}&Case study&Medical&A software for patient monitoring&\ON&Authors&Authors
                                \\ \textbf{Graydon et al.}  \cite{46_graydon2013}&Case study&Security&Security standards&\ON&Authors&Authors
                                \\ \textbf{Haley et al.}  \cite{54_haley2005}&Illustration&Software Engineering&Example HR system&&NA&Authors
                                \\ \textbf{Hawkins et al.}  \cite{15_hawkins2015}&Illustration&Model-Based Engineering&Cryptographic controller system&\ON&Authors&Authors
                                \\ \textbf{He et al.} \cite{4_he2012}&Case study&Medical&Lessons learned from security incidents, security standards, policies, and procedures&\ON&Authors&Authors
                                \\ \textbf{Masumoto et al.}  \cite{6_masumoto2013}&Experiment&Software Engineering&Commercial web application&\ON&Authors&Authors
                                \\ \textbf{Mohammadi et al.}  \cite{51_mohammadi2018}&Illustration&Medical&OPerational Trustworthiness Enabling Technologies (OPTET) research project &\ON&Authors&Authors
                                \\ \textbf{Netkachova et al.}  \cite{14_netkachova2015}&Illustration&Aviation&A security gateway data-flow controller&\ON&Authors*&Authors*
                                \\ \textbf{Netkachova et al. (2)}  \cite{30_netkachova2014}&Case study&Critical infrastructure&Electrical power system&\ON&Authors*&Authors*
                                \\ \textbf{Patu et al. (2)}  \cite{41_patu2013}&Illustration&Networking&Research e-learning project&\ON&Authors&Authors
                                \\ \textbf{Poreddy et al.}  \cite{26_poreddy2011}&Illustration&Aviation&Avionic mission control computer system&\ON&Authors&Authors
                                \\ \textbf{Ray et al.}  \cite{16_ray2015}&Illustration&Medical&A medical cyber-physical system for pumping insulin&\ON&Authors&Authors
                                \\ \textbf{Rodes et al.}  \cite{25_rodes2014}&Illustration&Security&Example scenario with confidence properties measurement&\ON&Authors*&Authors*
                                \\ \textbf{Shortt et al.} \cite{36_shortt2015} &Illustration&Software Engineering&Java-based open source library (Crawler4J)&&NA&Authors
                                \\ \textbf{Sklyar et al. (3)}  \cite{28_sklyar2017}&Illustration&SafSec&Requirements derived from safety and security standard&\ON&Authors&Authors
                                \\ \textbf{Sklyar et al. (4)}  \cite{31_sklyar2019}&Illustration&Medical&Example medical system&\ON&Authors&Authors
                                \\ \textbf{Sljivo et al.}  \cite{23_sljivo2016}&Illustration&Aviation&Wheel breaking system&\ON&Authors&Authors
                                \\ \textbf{Strielkina et al.}  \cite{22_strielkina2018}&Illustration&Medical&Healthcare IoT system&&NA&Authors
                                \\ \textbf{Tippenhauer et al.}  \cite{7_tippenhauer2014}&Illustration&Electrical&An electrical power grid use case&\ON&Authors&Authors
                                \\ \textbf{Vivas et al.}  \cite{34_vivas2011}&Illustration&Software Engineering&The research project PICOS (Privacy and Identity Management for Community Services)&\ON&Authors&Authors
                                \\ \textbf{Xu et al.} \cite{20_xu2017}&Case study&Software Engineering&IM server&\ON&Authors*&Authors*
            \\ \bottomrule
	\end{tabular}}
\end{table*}

%% file: Discussion.tex
\section{Discussion}\label{sec:disc}

Our findings show that the area of security assurance cases has not yet reached the same level of maturity as their safety counterpart has. In the following sub-sections, we will list the various reasons for that.

We also realised through our study that there is agreement about how security assurance cases are constructed in the literature. However, this agreement is not expressed in sufficient level of detail in any one paper yet. Therefore, we have synthesised the existing knowledge into a generic workflow for the construction of SAC that is presented in Section~\ref{disc:workflow}.

Main insights into the body of knowledge:
\begin{itemize}
    \item Many good motivations and usage scenarios, but not reflected.
    \item Room for creativity, and making use of the knowledge in safety in the approaches.
    \item Room for improvement on the support (tools and patterns)
    \item Lack of industrial validation of both approaches and usage scenarios
\end{itemize}
Synthesis is workflow which combines the different ideas and approaches from the body of knowledge
\subsection{Motivation and usage scenarios}
We were able to identify many motivations and usage scenarios from literature for using SAC. However, our impression is that these are on a high level and lack detailed studies to show how realistic and applicable they are.
For example, many papers motivate SAC for compliance with regulation and standards. However, they do not pinpoint the specific requirements for using SAC in these regulation and standards. It is hard to determine whether SACs are explicitly required, or if it is recommended as a way to create a structured argument for security without having previous knowledge of the specific regulation or standard.

\subsection{Approaches}
Most of the reported approaches to create complete assurance cases (including the argumentation and evidence) are top-down. That is, they start building from a top claim, all the way to sub-claims and evidence. Yamamoto \cite{49_yamamoto2015} who made an evaluation of architecture based on security and safety claims used a bottom-up technique starting from the evidence all the way up to the strategy and first claim.
No approach suggests a bottom-up approach starting from the evidence, i.e., building up from the existing evidence to form the argumentation and conclude what can be claimed about the system's security. This approach can be useful if the SAC is built for an already existing system. The evaluation can be then done between the ``what we can claim'', and the ``what goals and requirements does the system have'' questions. 
However, some of the selected studies used a hybrid approach where the SACs were still created top-down, but some parts of them were built using bottom-up techniques. An examples of a bottom-up technique used in literature is FMEA \cite{fmea} used in two studies \cite{15_hawkins2015,43_finnegan2013}.

Many of the approaches presented in literature treat security and safety cases as the same artifacts, e.g., \cite{11_chindamaikul2014,46_graydon2013,15_hawkins2015,23_sljivo2016,3_goodger2012,35_fung2018,9_ankum2005,24_gacek2014,31_sklyar2019,28_sklyar2017}.  We believe that since assurance cases in general are mature in the safety domain and have been used for a long time, it is natural to consider the gained knowledge and transferring it into other domains, such as the security domain. However, this knowledge transfer has to take into consideration the differences between safety and security e.g. in terms of field maturity and nature. Although we decided to include these studies in this SLR, we still think that their applicability in security needs further investigation taking into consideration the differences between the domains. Alexander et al. \cite{53_alexander2011} have a discussion on the differences between safety and security both from theoretical, and practical aspects. Other studies combine security and safety assurance by creating combined arguments or security-informed safety arguments \cite{48_taguchi2014,19_netkachova2016,2_cockram2007,14_netkachova2015}.

Another observation we made is the lack of quality assurance in the approaches. The approaches to create the arguments for example do not discuss the ability of the approach to assure the quality of the argumentation when it comes to its completeness, i.e., does the argumentation cover all and only the relevant security claims of the system? When it comes to the evidence creation approaches, there is a lack of quality assurance of confidence of the evidence, i.e., how confident are we that the provided evidence are enough to justify the claims associated with it?

\subsection{Validation}
Our results show a clear domination of illustration as the chosen type of validation in the included literature, which is an indicator of lack of industrial involvement. This explains why the majority of data used for validation are from research projects and example systems. Furthermore, the creation and validation of SAC in literature is mainly done by the authors of the studies, except for a few cases.
We believe that this contributed heavily to the lack of studies that address challenges and drawbacks of applying SACs in an industrial context.

It is fair to say that based on literature, applying SAC is not a trivial task, and doing so in an industrial context increases the complexity drastically. However, most of the reported motivations and usage scenarios are applicable in industry. Hence, there is a clear gap between research and industry, which needs to be addressed and closed.

\subsection{Tools}
Despite the fact that many of the selected papers included an example of a SAC, only a few reported on supporting tools for creating these cases. 
By extracting tool support information, it became clear that there is no set of commonly used tools for performing security cases tasks, such as creation documentation and maintenance. In fact
only one tool, ASCE~\cite{tool_adrelard}, has been reported to be used in more than one study. Moreover, there is a lack of tools addressing complex issues in SAC, such as claims dependencies, traceability, and quality control, e.g., evidence coverage. We did identify tools that address some of these matters \cite{35_fung2018,36_shortt2015}, but we believe there is a need for validating these in industry and specifically for security cases.

\subsection{SAC creation workflow}
\label{disc:workflow}
\begin{figure*}
\begin{center}
  \includegraphics[width=\textwidth]{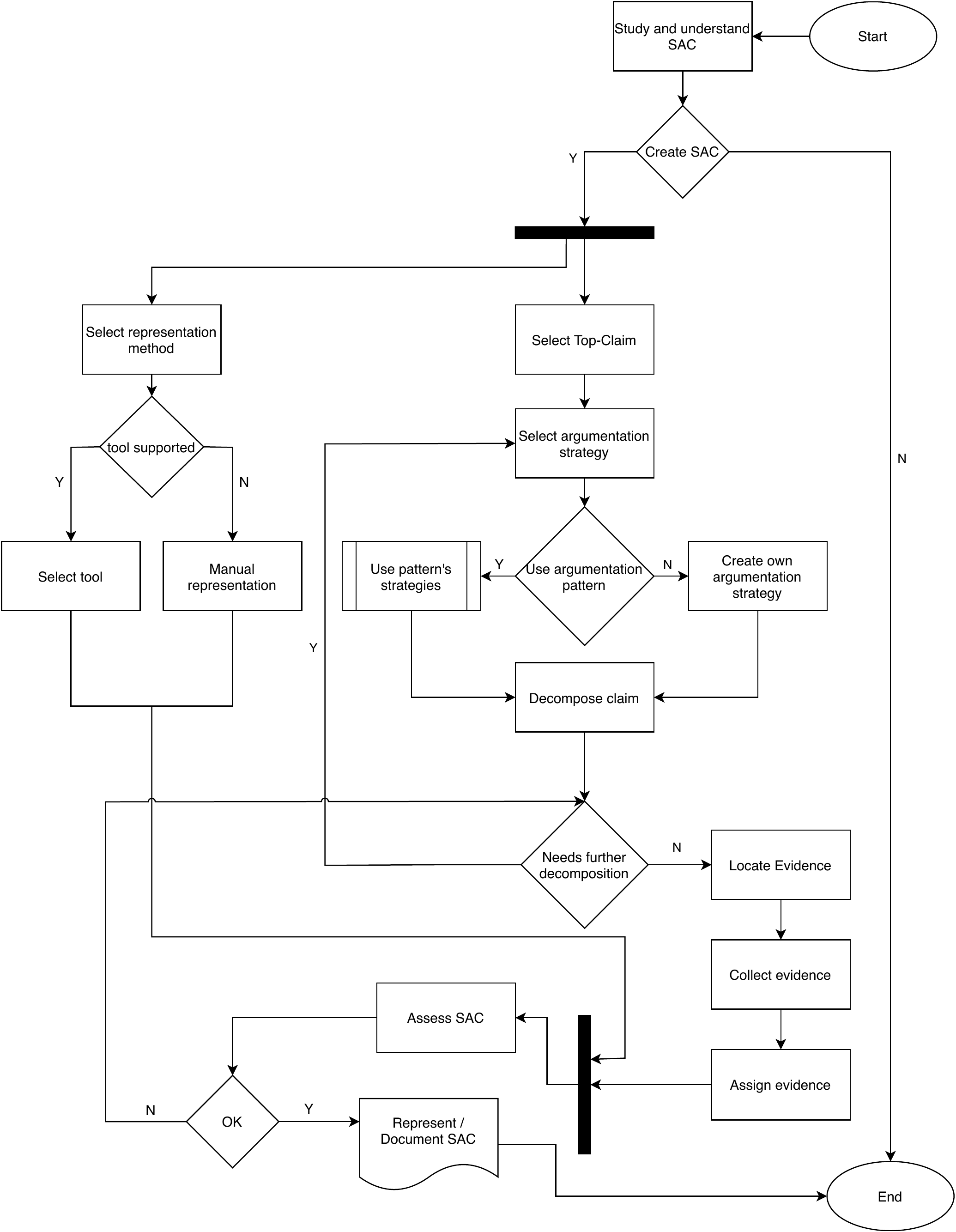}
\end{center}
        \caption{Flowchart of SAC creation} 
        \label{fig:flowchart}
\end{figure*}

Based on the results of this systematic literature review, we have found that the outcomes of the literature fall in one or more parts of the workflow depicted in Figure~\ref{fig:flowchart}.
Our flow diagram follows the top-down approach used in most reported approaches in literature.

Ankrum et. al. \cite{9_ankum2005} created a non-deterministic workflow for developing a structured assurance case. However, the proposed flow does not include anything related to tools or patterns usage. It does not consider the preliminary stage of considering a SAC either.
Cyra and Gorski \cite{1_cyra2007} present the life-cycle, derivation procedure, and application process for a trust case template. All these artifacts, however, build on the argumentation strategy being derived from a standard, which is not always the case.

There are five main blocks in the workflow. We will list and describe them in the remainder of this subsection. Additionally, we recommend papers to read which focus on aspects related to the individual blocks.

\begin{figure}
\begin{center}
  \includegraphics[width=\linewidth]{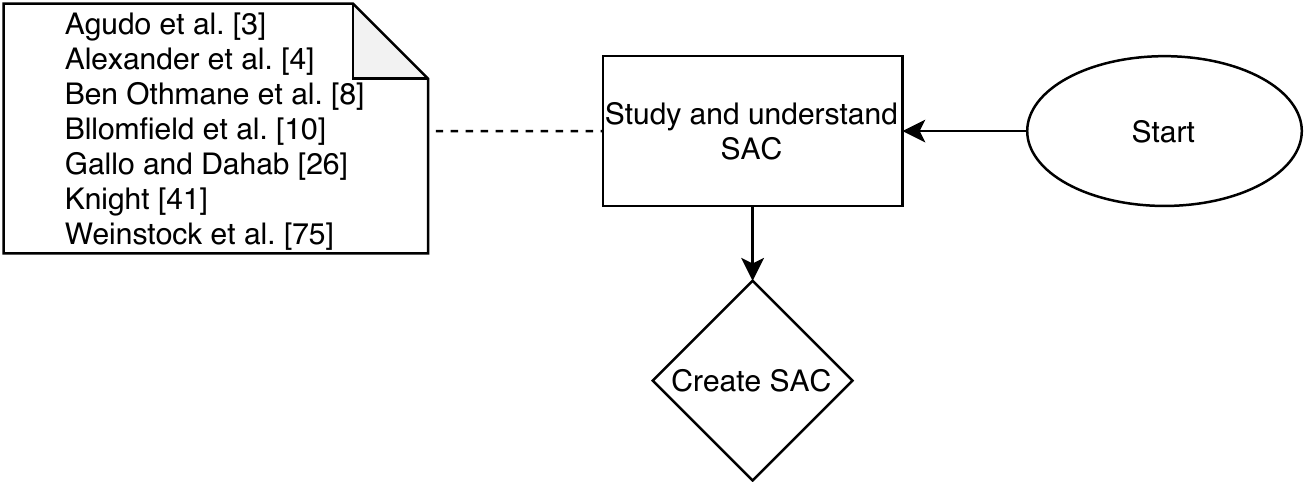}
\end{center}
        \caption{Corresponding entities and recommended studies for "Study and understand SAC" block} 
        \label{fig:flowBlock1}
\end{figure}

Study and understand SAC: Building SACs is not trivial. It requires a lot of work and dedication. Hence, before going ahead and creating them, it is important to understand what they are and what they can be used for. This step includes studying the structure of SACs, their benefits, what needs to be in place to create them, and their potential usage scenarios, e.g., standards and regulation compliance. Figure \ref{fig:flowBlock1} shows the corresponding entity in the workflow with the recommended papers, which are~\cite{53_alexander2011,52_weinstock2007,50_knight2015,38_agudo2009,bloomfield2010,37_gallo2015,10_othmane2014}.
\\ 

\begin{figure}
\begin{center}
  \includegraphics[width=\linewidth]{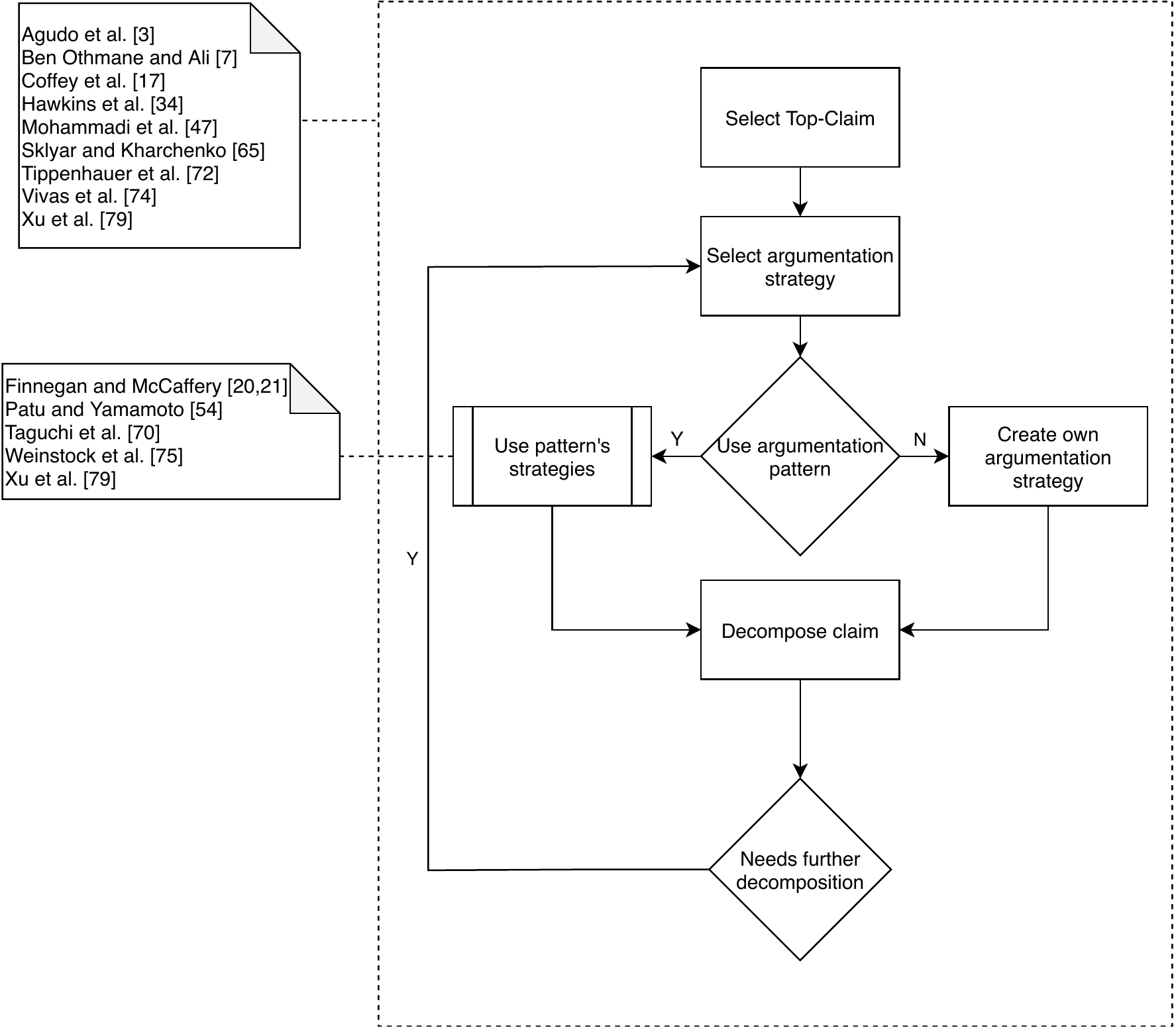}
\end{center}
        \caption{Corresponding entities and recommended studies for the Argumentation block} 
        \label{fig:flowBlock2}
\end{figure}

Argumentation: This block, which is depicted in Figure~\ref{fig:flowBlock2}, includes selecting the top claim to achieve, and the strategy to decompose this claim into sub-claims. This is a very important step, as selecting an argumentation strategy decides to a big extent what activities are needed to complete the SAC. For example, if a strategy where the decomposition is based on vulnerabilities is adopted, a vulnerability analysis of the system in question has to be conducted. The papers we recommend for this block are~\cite{34_vivas2011,20_xu2017,51_mohammadi2018,7_tippenhauer2014,15_hawkins2015,40_coffey2014,38_agudo2009,18_othmane2016}
A sub-block of the argumentation is the usage of patterns. Patterns help the creators of SACs to save time and effort by using pre-defined and proven structures. The creators could, however, decide not to use a pattern, and create their own unique structure if the situation requires that.
Creating a pattern is done based on the knowledge gathered while creating SACs. It is outside the scope of this workflow. However, this is discussed in the recommended papers~\cite{52_weinstock2007,48_taguchi2014,5_patu2013,8_finnegan2014,45_finnegan2014}.

\begin{figure}
\begin{center}
  \includegraphics[width=\linewidth]{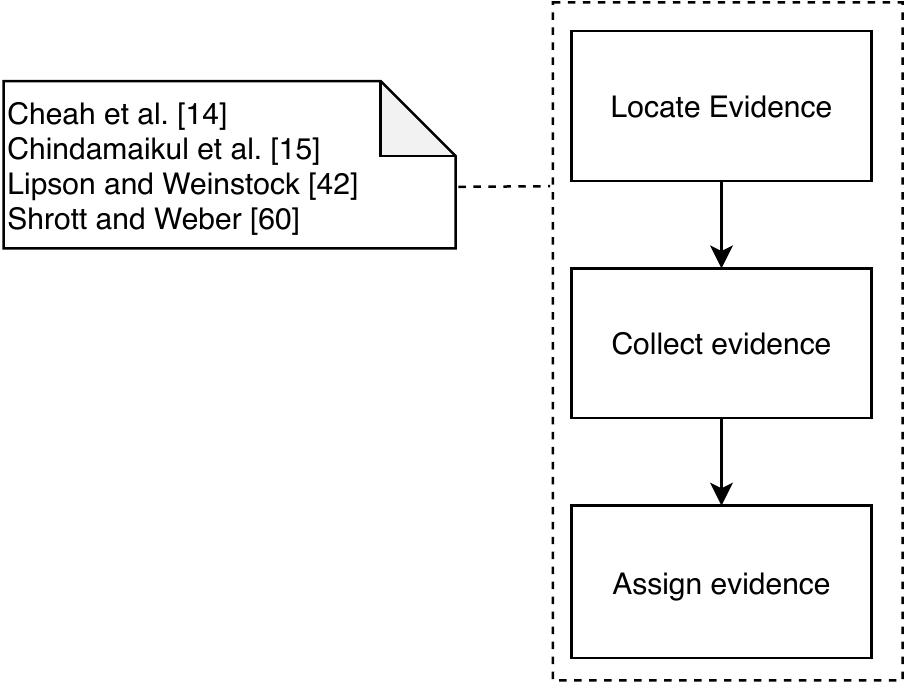}
\end{center}
        \caption{Corresponding entities and recommended studies for the Evidence block} 
        \label{fig:flowBlock3}
\end{figure}

Evidence: This block is shown in Figure~\ref{fig:flowBlock3}, and includes locating, collecting, and assigning evidence to the claims of the SAC. 
In some cases, the evidence is not present when the SAC is being built; hence, they need to be created. In our workflow, this would be a part of the collect evidence activity. Moreover, these activities might be done in an iterative manner. We consider the iterations to include an assessment of the SAC, e.g., to determine whether a claim needs extra evidence to reach a certain confidence level. For this block, we recommend these papers~\cite{56_lipson2008,11_chindamaikul2014,39_cheah2018,36_shortt2015}.

\begin{figure}
\begin{center}
  \includegraphics[width=\linewidth]{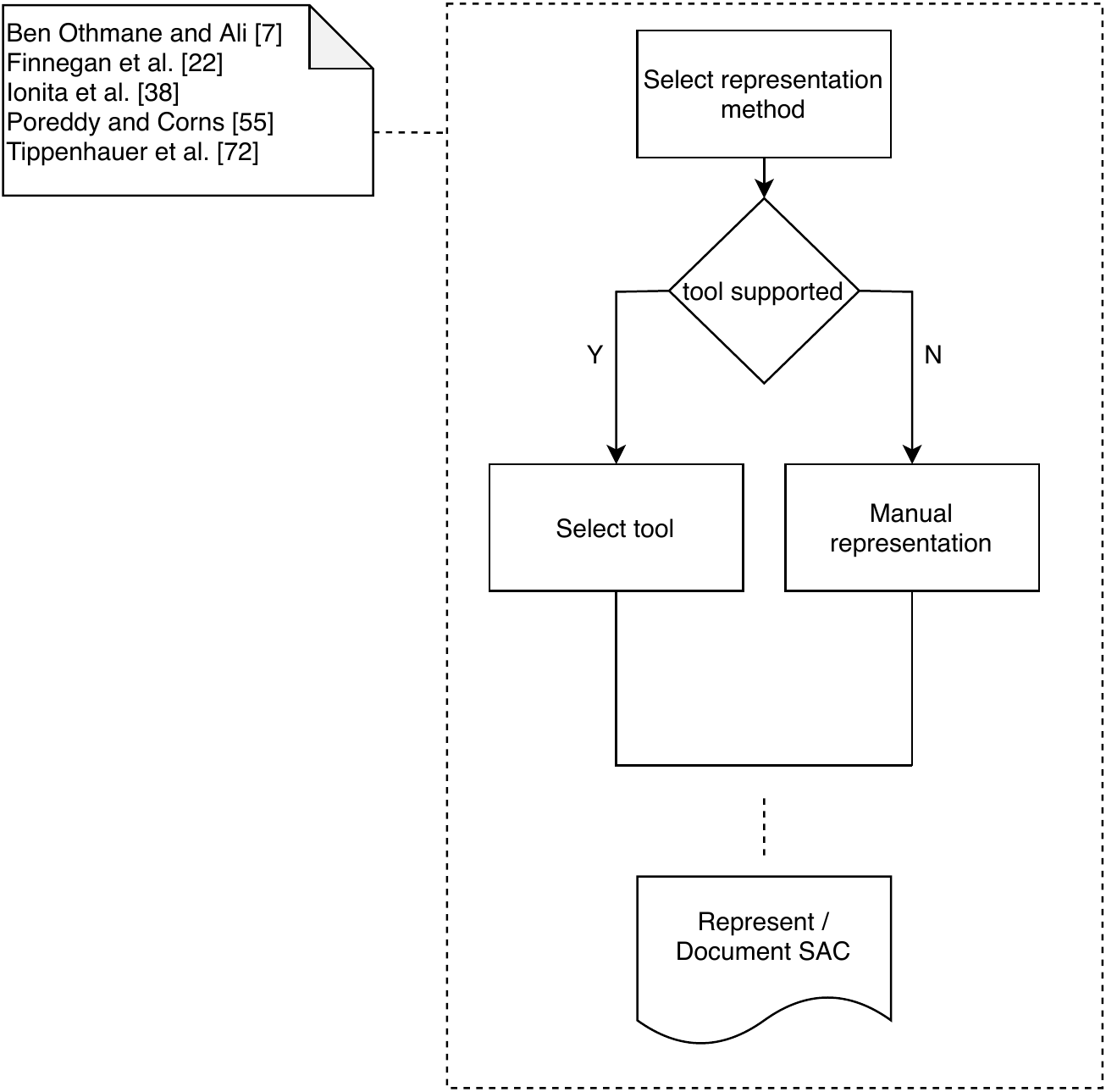}
\end{center}
        \caption{Corresponding entities and recommended studies for the Documenting block} 
        \label{fig:flowBlock4}
\end{figure}

Documenting: This block is depicted in Figure~\ref{fig:flowBlock4}. It includes making a decision of whether or not to use a tool for modelling the argument and documenting the SAC. If a tool is used, then the notation to be used is limited to the one/s supported by the tool. If a manual representation is to be done, then the creators will have the freedom to use an existing notation, extend one, or even create their own. We recommend the papers~\cite{47_ionita2017,7_tippenhauer2014,18_othmane2016,26_poreddy2011} for this block.

\begin{figure}
\begin{center}
  \includegraphics[width=\linewidth]{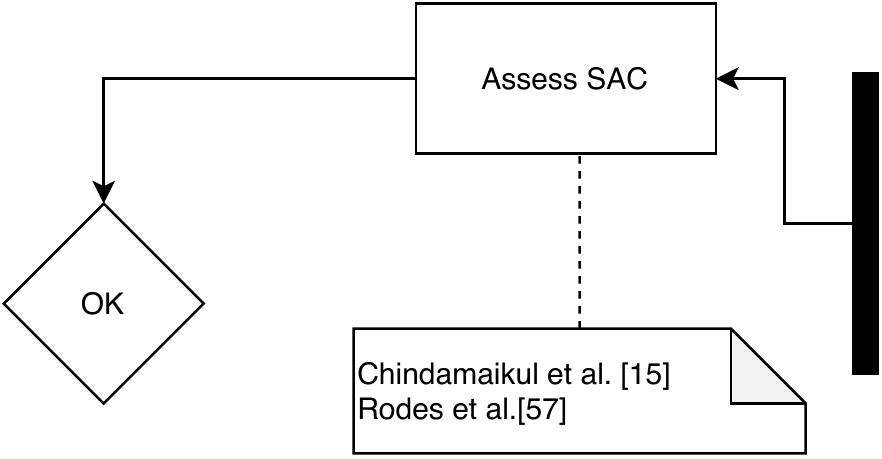}
\end{center}
        \caption{Corresponding entities and recommended studies for the Validation block} 
        \label{fig:flowBlock5}
\end{figure}

Assessment: This block is shown in Figure~\ref{fig:flowBlock5} and focuses on assessing SACs. This is done to check the quality of the created SAC, and determine whether it is sufficient, or needs additional work. Assessment starts after the claims have been identified and the evidence is assigned to the corresponding claims. The result of this step might require the creators of the SAC to go back to the point where they assess a claim and make a decision whether or not to further decompose it or assign evidence to it.
Since there is a lack of studies that focus on quality assurance of SAC, we have recommended studies which include some metrics to help assessing SACs~\cite{25_rodes2014,11_chindamaikul2014}.

%% file: ThreatsToValidity.tex
\section{Validity Threats}\label{sec:valTh}

In this study, we consider the internal and external categories of validity threats as defined in \cite{ttv1}, and described in \cite{ttv3,kh2007guidelines}. 
The work of conducting the review was done by one researcher. This means that applying the inclusion / exclusion criteria in each of the four filtering rounds was done by one person. This imposes a risk of subjectivity, as well as a risk of missing results, which might have affected the internal validity of this study. To mitigate this, a preliminary list of known good papers was manually created and used for a sanity check of the selected and included papers. Additionally, a quality control was performed periodically by the other authors to check the included and excluded studies.

Restricting our search to three digital libraries could have increased the probability of the risk of missing relevant studies. This was mitigated by performing the snowballing search to search for papers that are not necessarily included in the databases of the three considered libraries.

Another threat to validity is publication bias \cite{kh2007guidelines}. This is due to the fact that studies with positive results are more likely to get published than those with negative results. This could compromise the conclusion validity of this SLR, as in our case we did not find any study that is, e.g., against using SAC, or which reported a failed validation of its outcome. To mitigate this, we have scanned gray literature as part of the snowballing search. Additionally, in our preliminary work, we scanned the proceedings of eight conferences in the past five years.

External validity depends on the internal validity of the SLR \cite{kh2007guidelines}, as well as the external validity of the selected studies. We did scan gray literature to mitigate publication bias, but we excluded studies that are under 3 pages, and old studies as exclusion criteria to mitigate the risk of including studies with high external validity threats.

When it comes to the reliability of the study, we believe that any researcher with access to the used libraries will be able to reproduce the study, and get similar results plus additional results for the studies which get published after the work of this SLR is done.


%% file: Conclusion.tex
\section{Conclusion and future work}\label{sec:con}

In this study, we conducted a systematic review of the literature on security assurance cases. We used three digital libraries as well as snowballing to find relevant studies. We included 51 studies as primary data points, and extracted the necessary data for the analysis. 
The main findings of our study show that many usage scenarios for SAC are mentioned, and that several approaches for creating them are discussed. However, there is a clear gap between the usage scenarios and approaches, on one side, and their applicability in real world, on the other side, as the provided validations and tool support are far from being sufficient to match the level of ambition. 
Based on the results of this systematic literature review, we created a workflow for working with SAC, which is a useful tool for practitioners and also provides a guideline on how to approach the study of the literature, i.e., which paper is relevant in each stage of the workflow.

Based on our results and findings, in the future we will be working to close the gap between research and industry when it comes to applying security assurance cases. We will be looking into exact needs and challenges for these cases in specific domains, e.g., automotive. We believe that introducing SAC in large organizations needs appropriate planning to, e.g., find suitable roles for different tasks related to SAC, and integrating with current activities and way of working. Hence, we see a potential direction of future work in that area. 

When it comes to the technical work, we believe that there is room for improvement in the approaches for SAC creation, especially when it comes to the evidence part. For instance, a possible future work direction is to look into ways to automatically locate, collect, and assign evidence to different claims.

Finally, we believe that quality assurance of SAC has not been addressed sufficiently in literature. As a future work, we will look into ways to ensure the completeness of a security case when it comes to the argumentation, as well as the confidence in how well the provided evidence justify these claims.